\documentclass[aps,prb,twocolumn]{revtex4}

\usepackage{graphicx}
\usepackage{dcolumn}
\usepackage{amsmath}
\usepackage{amssymb}
\usepackage{wasysym}
\usepackage{color}

\usepackage{mathtools}

\begin{document}

\title{A highly asymmetric nodal semimetal in bulk SmB$_6$
}

\author{N.~Harrison$^1$
}

\affiliation{$^1$Mail~Stop~E536,~Los~Alamos~National Labs.,Los~Alamos,~NM~ 87545
}
\date{\today}

\begin{abstract}
We show that novel low temperature properties of bulk SmB$_6$, including the sudden growth of the de~Haas-van~Alphen amplitude (and heat capacity) at millikelvin temperatures and a previously unreported linear-in-temperature bulk electrical conductivity at liquid helium temperatures, signal the presence of a highly asymmetric nodal semimetal. We show that a highly asymmetric nodal semimetal is {\it also} a predicted property of the Kondo lattice model (with dispersionless $f$-electron levels) in the presence of Sm vacancies or other defects. We show it can result from a topological transformation of the type recently considered by Shen and Fu, and eliminates the necessity of a neutral Fermi surface for explaining bulk dHvA oscillations in SmB$_6$.
\end{abstract}
\maketitle


There is growing interest in the suggestion that certain members of the family of Kondo insulating compounds exhibit a strongly correlated topological insulating state,\cite{dzero1,takimoto1,dzero2} with electrical conduction taking place predominantly via the surface at low temperatures.\cite{kim1,wolgast1} While the  discovery of the de~Haas-van~Alphen (dHvA) effect in SmB$_6$ was widely considered to confirm the existence of pristine topologically protected surface metallic states,\cite{li1,xiang1} the cleanliness of the surface states has been brought into question by the absence of Shubnikov-de~Haas oscillations in the surface-dominated resistance.\cite{wolgast2} In another set of experiments,\cite{tan1,hartstein1} dHvA oscillations are reported to have the characteristic magnetic field angular-dependence of a bulk three-dimensional Fermi surface, leading to speculation over the possibility of the dHvA originating from novel neutral quasiparticles.\cite{kagan1,chowdhury1,coleman1,baskaran1} 
Various alternative yet more conventional explanations for the dHvA oscillations have also been proposed.\cite{erten1,knolle1,zhang1,shen1,ram1} A common feature of all the proposed models, however, whether based on neutral or conventional quasiparticles, is that they appear unable to account the sudden growth of the dHvA amplitude at millikelvin temperatures originating from a three-dimensional Fermi surface.\cite{hartstein1}


\begin{figure}[!!!!!!!htbp]
\centering 
\includegraphics*[width=.42\textwidth]{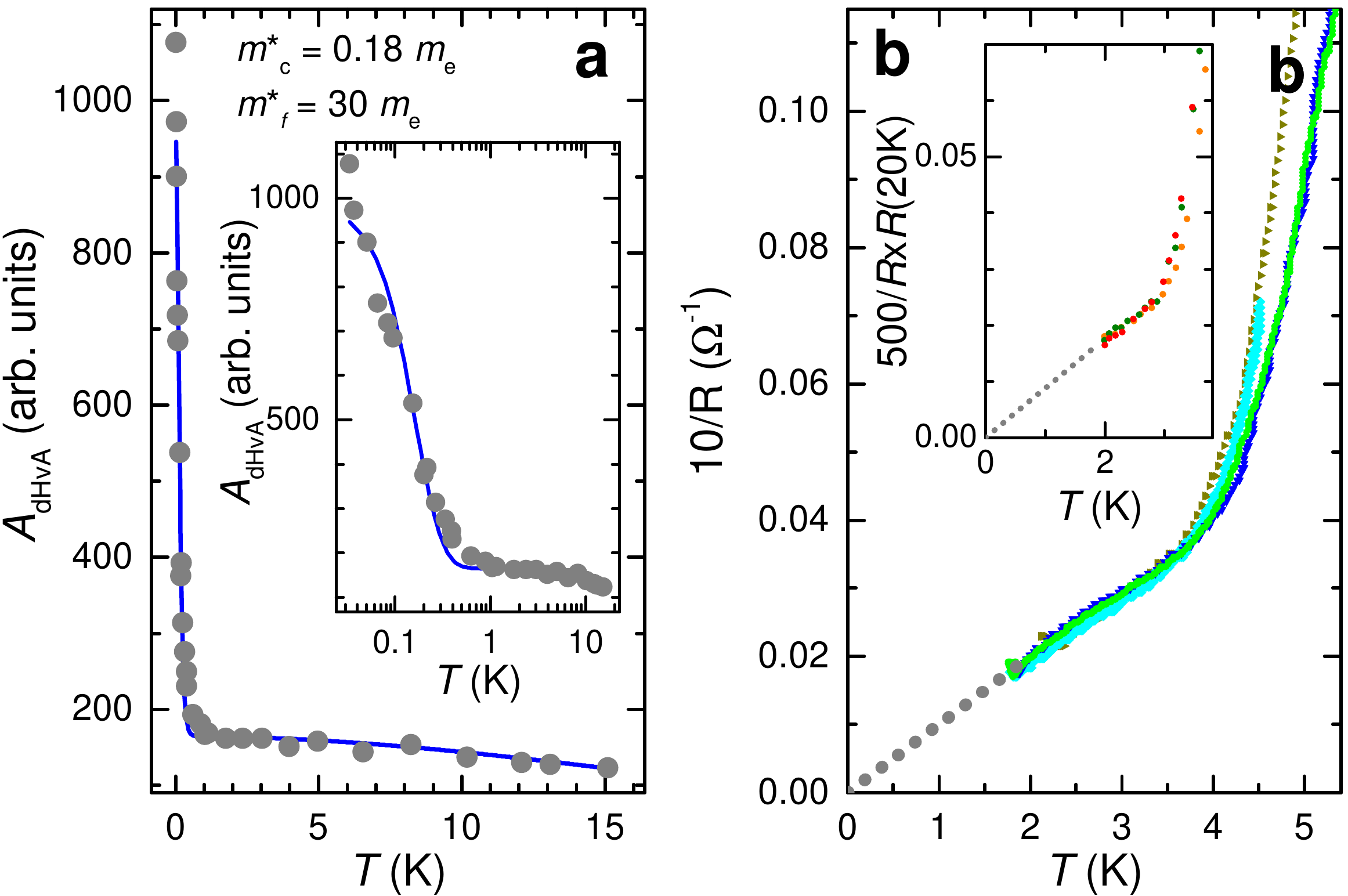}
\caption{({\bf a}), Temperature-dependent amplitude of the 330~T dHvA frequency in SmB$_6$ (black circles) together with a fit (blue line) to $A_{\rm dHvA}(T)=A_{\rm c}R_{T,{\rm c}}+A_fR_{T,f}$, in which $R_{T,{\rm c}}=X_{\rm c}/\sinh X_{\rm c}$ and $R_{T,f}=X_f/\sinh X_f$ are the thermal damping factors, $X_{\rm c}=2\pi^2m^\ast_{\rm c}k_{\rm B}T/ \hbar eB$ and $X_f=2\pi^2m^\ast_f k_{\rm B}T/ \hbar eB$, showing it to stem from the superposition of conduction electron-like and $f$-electron-like channels with effective masses $m^\ast_{\rm c}$ and $m^\ast_f$, respectively.  The inset shows the same fit with a logarithmic $T$ axis. ({\bf b}) Collapsed curves of the low $T$ region of the bulk conductance of SmB$_6$ inferred from Fig.~\ref{resistance}c, after subtracting the surface contribution $\sigma_{\rm surf}$ from each curve, with a dotted line extending to $T=0$ added as a guide to the eye. The inset shows similar collapsed curves from Fig.~\ref{resistance}b.}
\label{mass}
\end{figure}

In this paper, we present arguments for a highly asymmetric nodal semimetal existing over certain regions of momentum-space in bulk SmB$_6$, where the node is pinned to the unhybridized $f$-level. Our proposal is motivated by two recent experimental observations. The first is our finding that the dHvA effect\cite{hartstein1} in SmB$_6$ is consistent with two channels of identical frequency and similar mobility, but vastly different effective masses. One of the channels is of light conduction electron character while the other is of heavy $f$-electron character (see Fig.~\ref{mass}a). The second experimental observation is that the nonsaturating behavior of the resistivity plateau at liquid helium temperatures is caused by a bulk linear-in-temperature $T$ contribution to the electrical conductivity (see Fig.~\ref{mass}b). We attribute the linear-in-$T$ conductivity, which is evident in published data from both floating zone\cite{ciomaga1} and flux growth samples,\cite{syers1,wakeham1} to the thermal activation of uniform mobility carriers across the node. We propose the nodal semimetal to originate from defects in the crystalline lattice, such as Sm vacancies.\cite{valentine1}

\begin{figure}[!!!!!!!htbp]s
\centering 
\includegraphics*[width=.42\textwidth]{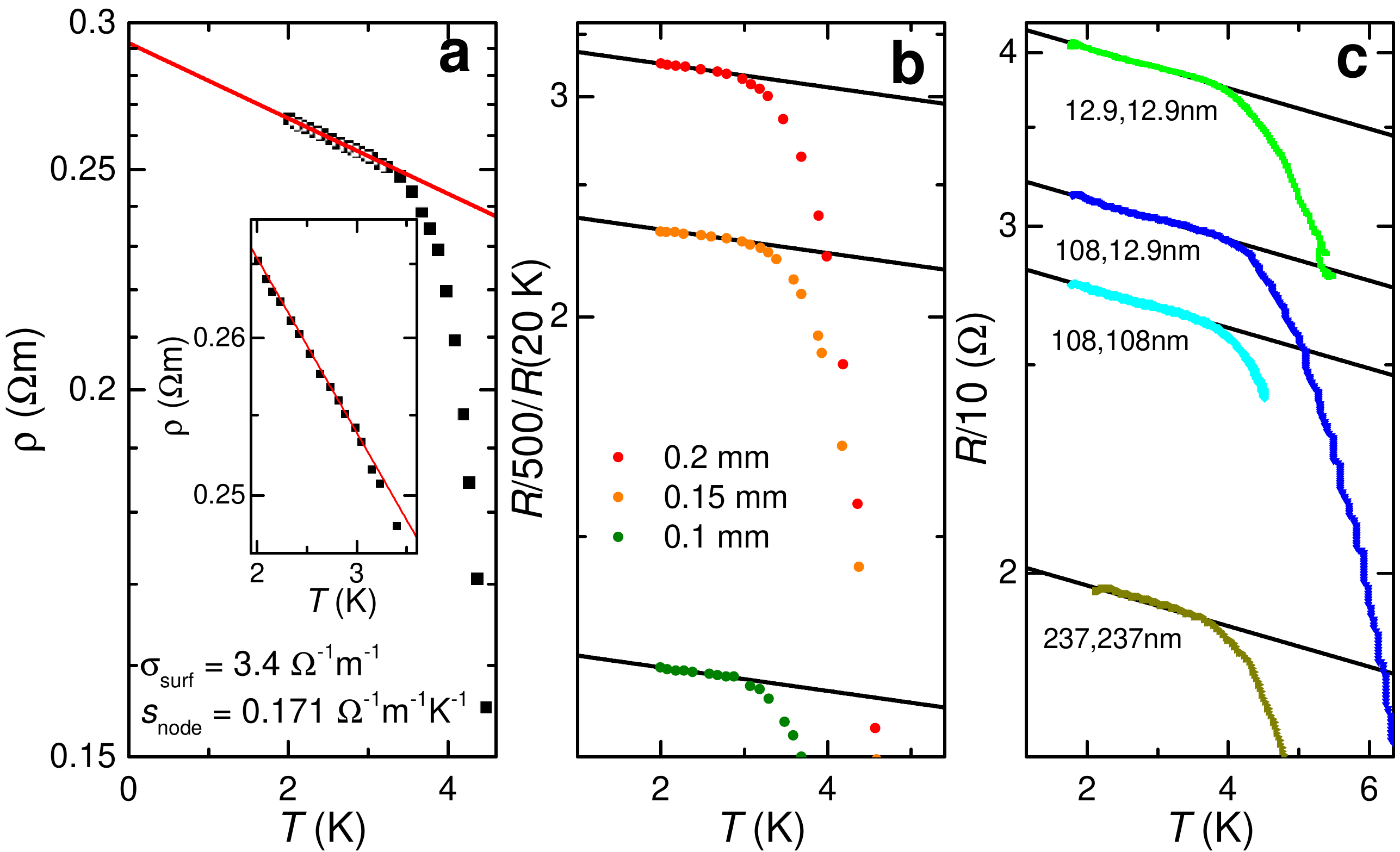}
\caption{({\bf a}), $T$-dependence of the electrical resistivity of SmB$_6$ grown using the floating zone technique\cite{ciomaga1} together with a fit to Equation~(\ref{resistancefit}) below $\approx$~4~K. ({\bf b}), $T$-dependent resistance (with reciprocal scaling) of flux grown SmB$_6$ crystals of different thicknesses rescaled so that their bulk-dominated resistances are the same at 20~K,\cite{syers1} showing that $s_{\rm node}T$ is invariant to thickness, indicating it to be of bulk origin. The black lines are given by Equation~(\ref{resistancefit}) in which $\sigma_{\rm surf}$ is adjusted to accommodate different sample thicknesses while $s_{\rm node}$ is held constant. ({\bf c}), Similar plot in which the surface of flux grown SmB$_6$ is radiation damaged to different depths (numbers shown for two faces),\cite{wakeham1} revealing that the surface contribution $\sigma_{\rm surf}$ is significantly impacted while $s_{\rm node}T$ is invariant, again indicating $s_{\rm node}T$ to be of bulk origin.}
\label{resistance}
\end{figure}


The experimental evidence for a linear-in-$T$ bulk contribution to the electrical resistivity is presented in Figs.~\ref{mass}b and \ref{resistance}, where we have used a reciprocal resistivity scale in order to display the resistivity in a manner that is proportional to conductivity. Figure~\ref{resistance}a shows the resistivity of a floating zone grown sample\cite{ciomaga1} while Figs.~\ref{resistance}b and c show multiple curves of the resistance measured on flux grown samples where only the surface contribution changes between vertically separated curves. A change in the relative surface conduction is achieved in Fig.~\ref{resistance}b by progressively reducing the sample thickness and rescaling the data at 20~K (where it is bulk-dominated,\cite{syers1} therefore making the bulk contributions the same) while this is achieved in Fig.~\ref{resistance}c by progressively increasing the depth of surface radiation damage.\cite{wakeham1} 
The simple manner in which the curves are vertically offset means that the conductance is dominated by a surface contribution $\sigma_{\rm surf}$ that is largely independent of $T$, thus confirming the conclusions reached in Refs..\cite{syers1,wakeham1} 
The new observation we make here is that the slope within the plateau region is invariant to changes in the surface conductance, revealing it to be of bulk origin. The low temperature resistivity therefore has the approximate form 
\begin{equation}\label{resistancefit}
\rho(T)\approx\frac{1}{\sigma_{\rm surf}+s_{\rm node}T},
\end{equation}
which we verify  in Fig.~\ref{resistance} by performing fits (black lines) in which $s_{\rm node}$ is held constant within each of the panels (a, b and c).


We further show how a highly asymmetric nodal semimetal can be a predicted property of a Kondo lattice in the presence of Sm vacancies (or other defects), which have been reported to exist at high concentrations in floating zone growth samples.\cite{valentine1} In the classic Kondo lattice picture, the $f$-electron levels start out as being strictly dispersionless (i.e. $\varepsilon_f=0$) and acquire dispersion {\it only} upon hybridization with conduction bands.\cite{martin1} The immobility of the unhybridized $f$-electrons implies that elastic scattering is expected to result {\it exclusively} from interactions between conduction electrons and defects. The corresponding energy level broadening is therefore $\Gamma_0=\frac{\hbar |{\bf v}_0|}{\lambda}$, where $\lambda$ is a semiclassical mean free path and ${\bf v}_0=\hbar^{-1}\partial\varepsilon_{\bf k}/\partial k$ is the Fermi velocity of the unhybridized conduction electron band. We proceed to obtain asymmetric nodal semimetal under these considerations by adapting the treatment recently introduced by Shen and Fu\cite{shen1} to the Kondo lattice scenario. We neglect any contributions to the energy broadening that are common to both $\varepsilon_f$ and $\varepsilon_{\bf k}$.\cite{footnote1}

\begin{figure}[!!!!!!!!!!!!!!!!!!!!!!!htbp]
\centering 
\includegraphics*[width=.42\textwidth]{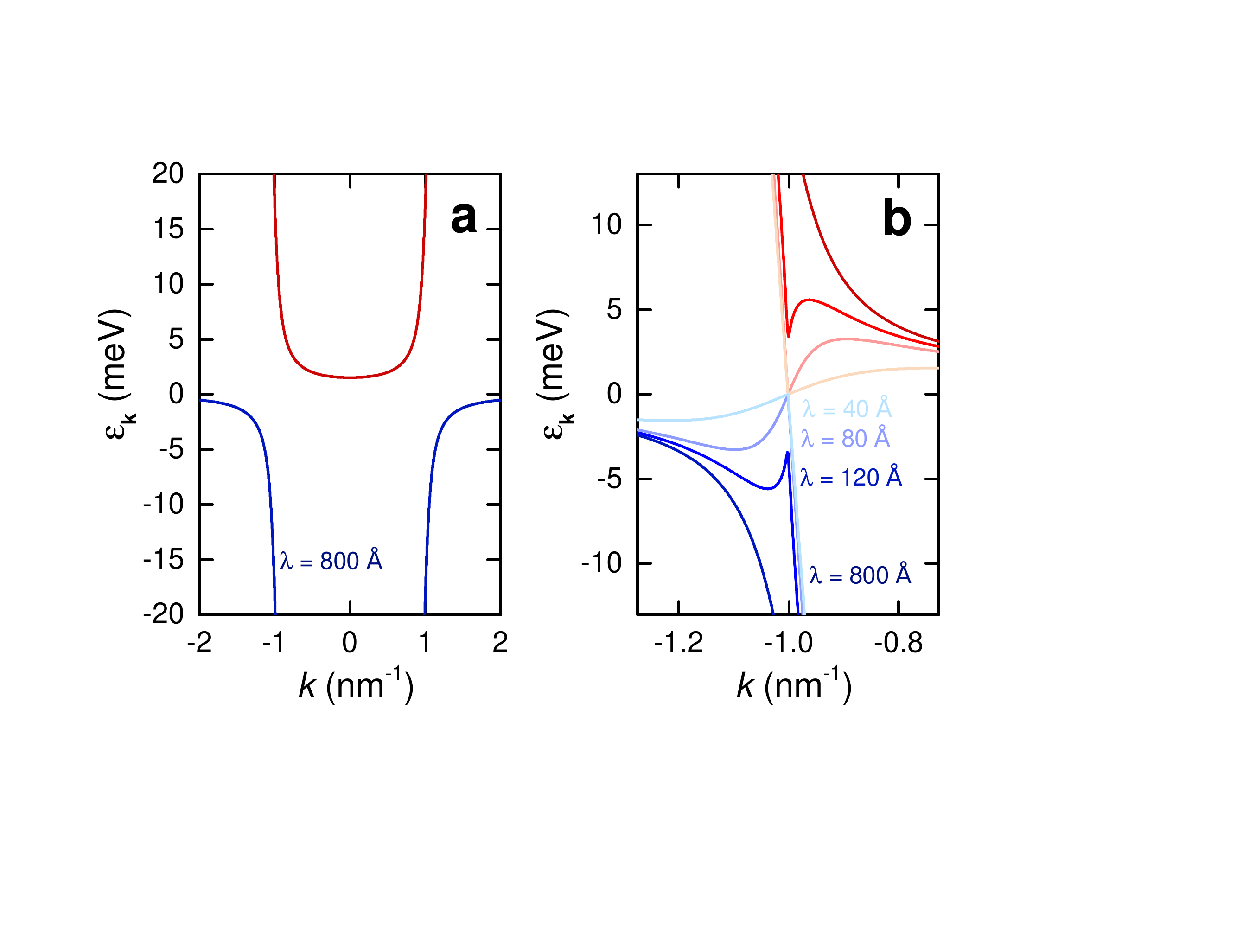}
\caption{({\bf a}), The real component of the reconstructed electronic dispersion $\varepsilon_{\bf k}$ for $F_0=$~330~T and $m^\ast_0\approx$~0.18~$m_{\rm e}$ (where $m_{\rm e}$ is the free electron mass), as for the small $\rho$ ellipsoids in SmB$_6$,\cite{hartstein1} for which we have $v_0\approx$~640,000~ms$^{-1}$, $k_0\approx$~1.00~$\times$~10$^9$m$^{-1}$ and $\varepsilon_0\approx$~21~meV. We use $\Delta\varepsilon=$~2~meV,\cite{shahrokhvand1} and $\lambda=$~800~\AA, which yields $\Gamma_0=\frac{\hbar |{\bf v}_0|}{\lambda}\approx$~5.3~meV. ({\bf b}), An expanded view of the electronic dispersion for the same $F$, $m^\ast$ and $\Delta\varepsilon$, but for different values of $\lambda$. $\lambda=$~800~\AA~corresponds to the situation in which $\Gamma_{\rm 0}\ll2V$, $\lambda=$~120~\AA~corresponds to the situation in which $\Gamma_{\rm 0}\lesssim2V$, $\lambda=$~80~\AA~corresponds to the situation in which $\Gamma_{\rm 0}\gtrsim2V$, and $\lambda=$~40~\AA~corresponds to the situation in which $\Gamma_{\rm 0}\gg2V$.}
\label{kondoinsulator}
\end{figure}

To simplify the derivation of the nodal semimetal, we approximate the unhybridized conduction electron band using a parabolic function
\begin{equation}\label{conductionband}
\varepsilon_{\bf k}=\frac{\hbar^2k^2}{2m^\ast_0}-\varepsilon_0,
\end{equation}
where $\varepsilon_0=2e\hbar F_0/m_0^\ast$ is defined in terms of the quantum oscillation frequency $F_0$ and the effective mass $m_0^\ast$ that have been extracted from fits to the dHvA oscillations at temperatures $T>$~1~K.\cite{hartstein1} 
After Shen and Fu,\cite{shen1} hybridization yields two bands
\begin{equation}\label{bands}
\varepsilon^{\pm}_{\bf k}-i\Gamma^\pm_{\bf k}=\frac{1}{2}(\varepsilon_{\bf k}+\varepsilon_f-i\Gamma_0)\pm\sqrt{\frac{1}{4}(\varepsilon_{\bf k}-\varepsilon_f-i\Gamma_0)^2+V^2},
\end{equation}
which can be separated into real and imaginary components, $\varepsilon^{\pm}_{\bf k}$ and $\Gamma^\pm_{\bf k}$, respectively. 
When $\Gamma_0\ll2V$, Equation (\ref{bands}) produces the classic dispersion of a Kondo insulator with a gap at the chemical potential (see Fig.~\ref{kondoinsulator}a). Here, we define the hybridization potential $V=\sqrt{(\Delta\varepsilon+\frac{\epsilon_0}{2})^2-\frac{\varepsilon_0^2}{4}}$ in terms of the gap $\Delta\varepsilon$ reported in low temperature transport and point contact spectroscopy measurements.\cite{shahrokhvand1,flachbart1} 

For $\Gamma_0\lesssim2V$, the imaginary term leads to a small reduction in the Kondo insulator gap (see Fig.~\ref{kondoinsulator}b). When $\Gamma_0$ and $2V$ become comparable in magnitude, however, a significant funnel-shaped depression in the gap emerges at $k=\pm k_0$ in Fig.~\ref{kondoinsulator}b. Once $\Gamma_0\geq 2V$, the system undergoes a topological transformation into a nodal semimetal.\cite{shen1} In this limit, $\Gamma_0$ overwhelms the Kondo insulator hybridization, causing the electronic dispersion to become gapless (see Fig.~\ref{kondoinsulator}b). In the vicinity of $k=\pm k_0$, the expansion of Equation~(\ref{bands}) into real and imaginary components is trivial. The real components can be be expanded relative to $|k|-k_0$, so that the electronic dispersion consists of two hybridized bands
\begin{eqnarray}\label{lineardispersion}
\tilde{\varepsilon}_{{\bf k},{\rm c}}=\tilde{v}_{\rm c}(|k|-k_0)+c_{\rm c}(|k|-k_0)^3+\dots\nonumber\\
\tilde{\varepsilon}_{{\bf k},f}=\tilde{v}_f(|k|-k_0)+c_f(|k|-k_0)^3+\dots
\end{eqnarray}
pinned to $\varepsilon_f$, where 
\begin{eqnarray}\label{velocities}
\tilde{v}_{\rm c}=\frac{\partial\varepsilon^+_{\bf k}}{\hbar\partial k}\bigg|_{k=k_0}=\frac{v_0}{2}\bigg[1+\frac{\Gamma_0}{\sqrt{\Gamma_0^2-4V^2}}\bigg]\nonumber\\
\tilde{v}_f=\frac{\partial\varepsilon^-_{\bf k}}{\hbar\partial k}\bigg|_{k=k_0}=\frac{v_0}{2}\bigg[1-\frac{\Gamma_0}{\sqrt{\Gamma_0^2-4V^2}}\bigg]
\end{eqnarray}
are the Fermi velocities at $k=\pm k_0$ (of opposite sign) and $c_{\rm c}$ and $c_{\rm f}$ are higher order terms. The subscripts ${\rm c}$ and $f$ refer to each of the bands being primarily of unhybridized conduction electron and $f$-electron character, respectively. For the imaginary components, we obtain
\begin{eqnarray}\label{gammas}
\tilde{\Gamma}_{\rm c}=\frac{\Gamma_0}{2}+\frac{\sqrt{\Gamma_0^2-4V^2}}{2}\nonumber\\
\tilde{\Gamma}_f=\frac{\Gamma_0}{2}-\frac{\sqrt{\Gamma_0^2-4V^2}}{2}
\end{eqnarray}
at $k=\pm k_0$. 

\begin{figure}[!!!!!!!!!!!!!!!!!!!!!!!htbp]
\centering 
\includegraphics*[width=.4\textwidth]{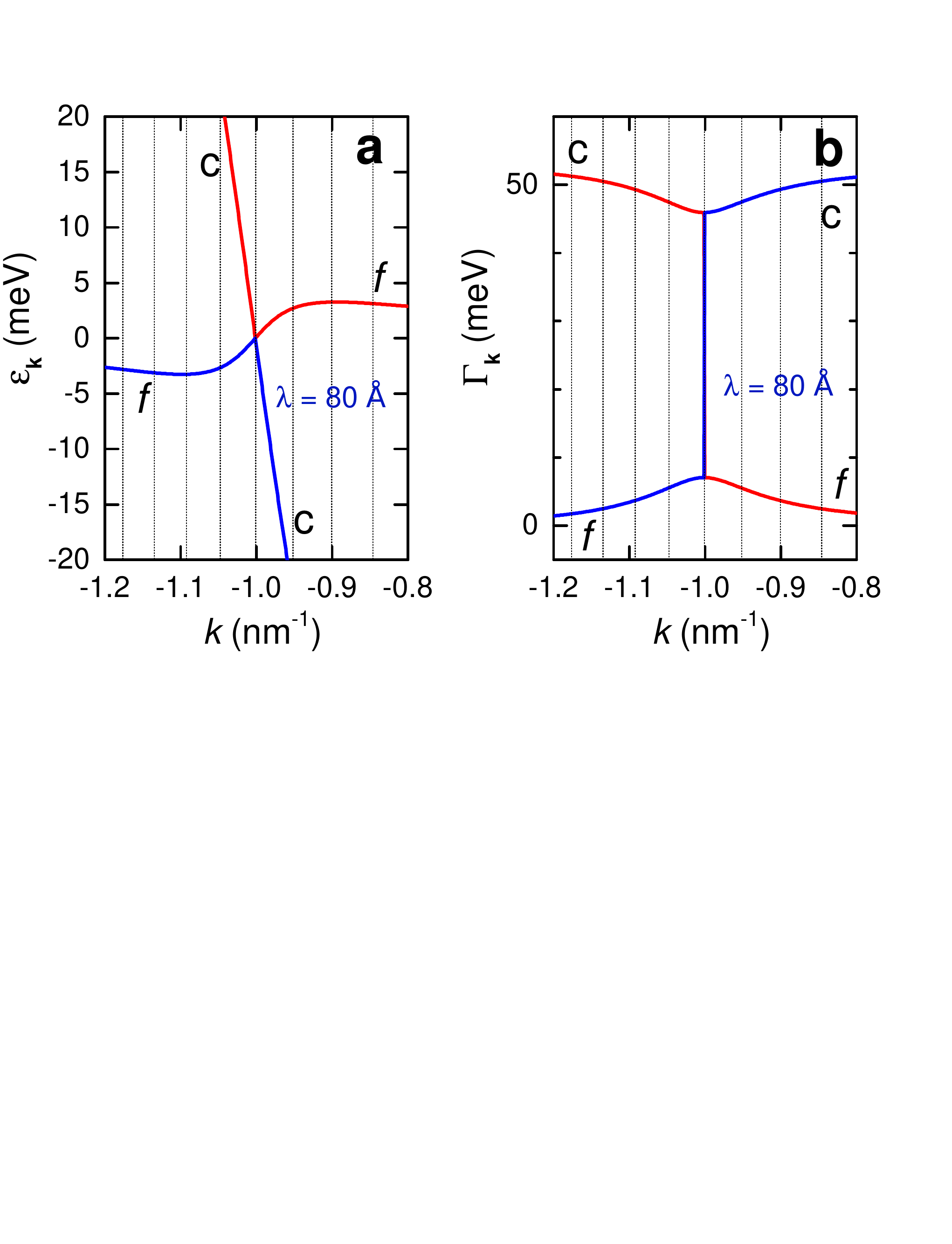}
\caption{({\bf a}), The real component of the reconstructed electronic dispersion $\varepsilon_{\bf k}$ for the small $\rho$ ellipsoids in SmB$_6$,\cite{hartstein1} using $\Delta\varepsilon=$~2~meV\cite{shahrokhvand1} and $\lambda=$~80~\AA, which yields $\Gamma_0=\frac{\hbar |{\bf v}_0|}{\lambda}\approx$~53~meV. ({\bf b}), The imaginary contribution for $\lambda=$~80~\AA. Vertical dotted lines indicate the values of $k$ (assuming circular geometry) at which Landau levels intersect $\varepsilon^\pm_{\bf k}$ and $\Gamma^\pm_{\bf k}$ at $B=$~31.4~T, given by all values of $k=\sqrt{\frac{2eB}{\hbar}(\nu+\frac{1}{2})}$ for which $\nu$ is an integer.}
\label{dispersion}
\end{figure}

A peculiar consequence of the dispersion and defect scattering originating solely from the unhybridized conduction electron band is that the mobility
\begin{equation}\label{mobility}
\mu^\ast=\frac{e\tau_f}{m^\ast_f}\equiv\frac{e\tau_{\rm c}}{m^\ast_{\rm c}}=\frac{ev_0}{k_0\sqrt{\Gamma_0^2-4V^2}}.
\end{equation}
is uniform with respect to the hybridized ${\rm c}$ and $f$ bands at $k=\pm k_0$, with the relaxation times and effective masses being given by $\tau_{\rm c}=\hbar/\tilde{\Gamma}_{\rm c}$ and $\tau_f=\hbar/\tilde{\Gamma}_f$, and $m^\ast_{\rm c}=\hbar k_0/|\tilde{v}_{\rm c}|$ and $m^\ast_f=\hbar k_0/|\tilde{v}_f|$, respectively. The uniform mobility has interesting implications for the behavior of SmB$_6$ at low temperatures. 

In the case of the dHvA effect, the different Fermi velocities in Fig.~\ref{dispersion} implies that the dHvA effect is the sum of two components with identical frequencies $F_0$ but very different effective masses, $m^\ast_{\rm c}$ and $m^\ast_f$, which become
$$m^\ast_{\rm c}\approx m_0~~~{\rm and}~~~m^\ast_f\approx m_0(\Gamma_0/V)^2$$ in the limit $\Gamma_0\gg2V$, meaning that $m^\ast_f\gg m^\ast_{\rm c}$. The heavier fitted effective mass in Fig.~\ref{mass}a corresponds to $\Gamma_0/V\approx$~13 at high magnetic fields, which suggests an upper bound of $V\sim$~4~meV for the hybridization potential in the vicinity of the small $\rho$ ellipsoids reported in SmB$_6$\cite{hartstein1,tan1} (given that $\Gamma_0$ must be of order 50~meV or less in order for dHvA oscillations to be observable). Since the exponent ($\pi/\mu^\ast B$) of the Dingle damping term (associated with scattering from defects) is proportional to the inverse mobility, the uniform mobility implies that the amplitudes of the two channels will be of similar order at low temperatures. A slightly larger magnitude for the $f$-electron branch may be the consequence of partially filled Landau levels away from the chemical potential with less broadening contributing to the dHvA effect, reflecting the fact that $\tilde{\Gamma}_{{\bf k},f}$ drops once $k\neq\pm k_0$ (see Fig.~\ref{dispersion}b) while $\tilde{\Gamma}_{{\bf k},{\rm c}}$ increases.

The observability of quantum oscillations requires $2V\leq\Gamma_0\lesssim\hbar\omega_{\rm c}$, which is easily realized for the small $\rho$ ellipsoids but not so easily realized for the large $\alpha$ ellipsoids owing to their smaller $\hbar\omega_{\rm c}$. On the other hand, dHvA oscillations from larger sections of Fermi surface have only been observed once.\cite{tan1} In the present model, the observation of $\alpha$ ellipsoid dHvA  becomes possible only under extremely fortuitous circumstances in which $\Gamma_0\approx2V$ in Equation~(\ref{mobility}).


In the case of the electrical transport, the conductivity is that of a nodal semimetal, which implies possible similarities to that of a line-node semimetal.\cite{carbotte1} In SmB$_6$, the excited carriers reside on a thin $k$-space film of area $S_k$ that encompasses the Fermi surface of the original unhybridized conduction electron band wherever the inequality $\Gamma_0\geq2V$ is satisfied. If the density of defects (e.g. Sm vacancies) is nonuniform, then the extreme sensitivity of the bulk conductivity to $\Gamma_0/V$ (becoming conducting for $\Gamma_0>2V$ and insulating for $\Gamma_0<2V$) implies that the total resistivity could be of a highly percolative nature. In the case of a uniform density, however, we can proceed to calculate the conductivity using a simple Drude picture. In this case, the low temperature bulk conductivity is given by
\begin{equation}\label{drude}
\sigma_{xx,{\rm bulk}}=ne\mu^\ast=s_{\rm node}T~~~{\rm and}~~~\sigma_{xy,{\rm bulk}}=0,
\end{equation}
where $\mu^\ast$ is the uniform mobility given by Equation~(\ref{mobility}),
\begin{equation}\label{number}
n=2\int_0^\infty{\rm f}^\prime(\varepsilon)D(\varepsilon)\varepsilon{\rm d}\varepsilon=\frac{\ln2~S_k}{4\pi^3\hbar v_0}\frac{\Gamma_0\sqrt{\Gamma^2_0-V^2}}{V^2}~k_{\rm B}T
\end{equation}
is the density of electrons plus holes thermally excited across the node, ${\rm f}^\prime(\varepsilon)$ is the derivative of the Fermi-Dirac function and 
\begin{equation}\label{density}
D(\varepsilon)=\frac{S_k}{4\pi^3\hbar}\bigg(\frac{1}{|v_{\rm c}|}+\frac{1}{|v_f|}\bigg)=\frac{S_k}{4\pi^3\hbar v_0}\frac{\Gamma_0\sqrt{\Gamma_0^2-V^2}}{V^2}
\end{equation}
is the electronic density-of-states (taking into consideration the asymmetry of the node). From this we obtain 
\begin{equation}\label{node}
s_{\rm node}=\frac{\ln2~e^2S_k}{4\pi^3\hbar k_0}\frac{\Gamma_0}{V^2}~k_{\rm B}. 
\end{equation}
The central prediction of the nodal semimetal, therefore, is that the bulk conductivity has a linear dependence on temperature -- a similar temperature-dependence having been predicted for the line-node semimetal\cite{carbotte1} -- and that the bulk Hall conductivity vanishes -- which appears to have already been indicated in sample thickness-dependent experiments.\cite{kim1} As $T$ is increased, the outer flanks of ${\rm f}^\prime(\varepsilon)$ eventually reach the bottom of the conduction band ($k=0$) and top of the valence band ($|k|>2$~nm$^{-1}$) in Fig.~\ref{kondoinsulator}a. The very large electronic density of states at these points is likely to be primarily responsible for the thermally activated contribution to the electrical conductivity [not included in Equation~(\ref{resistancefit})] that dominates the Kondo insulating state at high $T$. Since $\varepsilon_{\bf k}$ and $\varepsilon_f$ are vastly different in energy at $k=0$ and $|k|>2$~nm$^{-1}$, the high temperature thermally activated behavior of the Kondo insulator is not expected to be significantly impacted by the formation of a nodal semimetal. 

The fitted magnitude of the linear-in-$T$ conductivity is  found to be strongly sample-dependent in Fig.~\ref{resistance}, yielding $s_{\rm node}\approx$~0.17, 0.0028 and 0.43~$\Omega^{-1}$m$^{-1}$K$^{-1}$ for the samples in panels a, b and c, respectively. On considering this spread of values together with $\Gamma_0=$~53~meV and different values of the ratio $\Gamma_0/V$ ranging from 2 to 13, we find the fitted $s_{\rm node}$ in Fig.~\ref{resistance}a to be consistent with a $k$-space area $S_k$ ranging between 4~$\times$~10$^{14}$~m$^{-2}$ and 3~$\times$~10$^{18}$~m$^{-2}$. The upper end of this range is similar to the cross-sectional area of one of the $\rho$ ellipsoids in SmB$_6$. The highly elongated shape of the $\rho$ ellipsoids in SmB$_6$ implies that $v_0$ and hence $\Gamma_0$ are $\approx$~2.4 times smaller around the belly of the ellipsoids than at the tips, which means that the nodal semimetal will initially form around the belly. Angle-dependent dHvA $m^\ast_f$ measurements could be used to investigate how far such a nodal semimetal extends in $k$-space away from the belly regions. 


Heat capacity measurements have suggested that the low temperature heat capacity of SmB$_6$ increases with the concentration of Sm vacancies.\cite{valentine1} If this is confirmed to be a general trend, then it means that there should be a concomitant increase in $s_{\rm node}$ and, perhaps, also $A_{\rm dHvA}$ in Fig.~\ref{mass}a. In addition to producing a very heavy effective mass component, a very large $\Gamma_0$ causes the reconstructed $f$-band in Fig.~\ref{kondoinsulator}b to become very narrow, potentially providing an explanation for the upturn in the electronic heat capacity at low temperatures.\cite{hartstein1} Conversely, a nodal semimetal is not expected to occur in a perfectly stoichiometric defect-free sample,\cite{eo1} which could be verified by the vanishing of $s_{\rm node}$ in such samples. 

Dingle plots, in which the logarithm of the dHvA amplitude is plotted versus $1/B$,\cite{shoenberg1} provide a possible means for isolating the  mobilities of the heavy and light carriers. The slope of the Dingle plot is $-\pi/\mu^\ast$, which means that of the slopes of the $f$-electron-dominated Dingle plot at dilution fridge temperatures and the conduction electron-dominated Dingle plot at $T\gtrsim$~1~K should be the same.

In summary, we propose a highly asymmetric nodal semimetal to exist over certain regions of momentum-space of bulk SmB$_6$, which we show to be supported by recent experimental data. The first is that the dHvA amplitude\cite{hartstein1} is consistent with two channels of identical frequency and similar mobility, but vastly different effective masses, $m^\ast_{\rm c}\approx$~0.18~$m_{\rm e}$ and $m^\ast_f\approx$~30~$m_{\rm e}$. The second is a bulk linear-in-temperature $T$ contribution to the electrical conductivity evident in the nonsaturating behavior of the resistivity plateau at liquid helium temperatures.\cite{syers1,wakeham1,ciomaga1} We show that these observations can be qualitatively understood by considering the effect of lattice defects on the Kondo lattice model, to which we adapt a theoretical treatment recently developed by Shen and Fu.\cite{shen1} A strong possibility, therefore, is that the nodal semimetal is caused by defects in the crystalline lattice, such as those arising from Sm vacancies.\cite{valentine1} The existence of a nodal semimetal does not exclude other conventional quasiparticle origins of the dHvA over other regions of $k$-space.\cite{knolle1,erten1,zhang1,shen1} It does, however, cast doubt over the necessity of a neutral Fermi surface.\cite{hartstein1}

This work was supported by the US Department of Energy ``Science of 100 tesla" BES program. The author acknowledges insightful discussions with Arkhady Shekter, Ross McDonald and Priscila Rosa.


\begin{thebibliography}{99}
\bibitem{footnote1} Such a contribution survives the hybridization unaltered, thereby not contributing to the nodal semimetal.\cite{shen1}
\end{thebibliography}
\end{document}